\documentclass[english,reprint,amsmath,amssymb,aps,pra,superscriptaddress]{revtex4-1}
\usepackage[latin9]{inputenc}
\usepackage{amsmath}
\usepackage{graphicx}
\usepackage{esint}

\makeatletter
\usepackage{dcolumn}
\usepackage{bm}

\makeatother

\usepackage{babel}
\begin{document}

\title{Cooper problem in a cuprate lattice}

\author{Ali Sanayei}
\email{asanayei@physnet.uni-hamburg.de}

\address{Zentrum f{\"u}r Optische Quantentechnologien, Universit{\"a}t Hamburg,
Luruper Chaussee 149, D-22761 Hamburg, Germany}

\address{Institut f{\"u}r Laserphysik, Universit{\"a}t Hamburg, Luruper
Chaussee 149, D-22761 Hamburg, Germany}

\author{Ludwig Mathey}
\email{lmathey@physnet.uni-hamburg.de}

\address{Zentrum f{\"u}r Optische Quantentechnologien, Universit{\"a}t Hamburg,
Luruper Chaussee 149, D-22761 Hamburg, Germany}

\address{Institut f{\"u}r Laserphysik, Universit{\"a}t Hamburg, Luruper
Chaussee 149, D-22761 Hamburg, Germany}

\address{The Hamburg Centre for Ultrafast Imaging, Luruper Chaussee 149, D-22761
Hamburg, Germany}

\date{\today}
\begin{abstract}
We solve the Cooper problem in a cuprate lattice by utilizing a three-band
model. We determine the ground state of a Cooper pair for repulsive
on-site interactions, and demonstrate that the corresponding wave
function has an orbital $d_{x^{2}-y^{2}}$ symmetry. We discuss the
influence of next-nearest-neighbor tunneling on the Cooper pair solution,
in particular the necessity of next-nearest-neighbor tunneling for
having \emph{d}-wave pairs for hole-doped systems. We also propose
experimental signatures of the \emph{d}-wave Cooper pairs for a cold-atom
system in a cuprate lattice.
\end{abstract}
\maketitle

\section{introduction \label{Sec_I}}

\begin{figure}
\includegraphics[width=1\columnwidth]{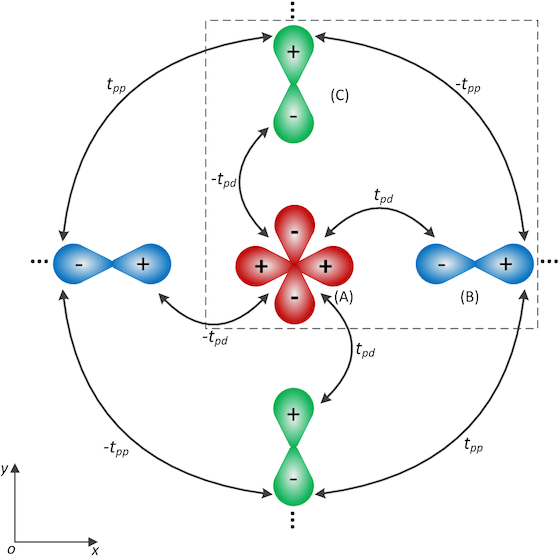}

\caption{Sketch of the two-dimensional cuprate lattice in real space. The unit
cell is shown by a dashed square, including a $d_{x^{2}-y^{2}}$ orbital
configuration on the A-site, a $p_{x}$ orbital on the B-site, and
a $p_{y}$ orbital on the C-site. The nearest-neighbor hopping is
shown by $t_{pd}$ and the next-nearest-neighbor hopping is shown
by $t_{pp}$.}

\label{Fig1}
\end{figure}

\begin{figure*}
(a)\includegraphics[width=0.4\textwidth]{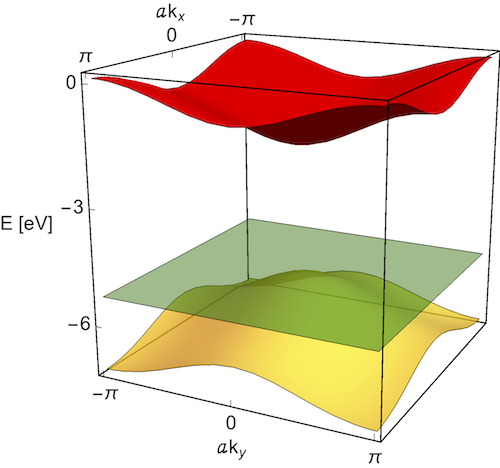}$\qquad$(b)\includegraphics[width=0.4\textwidth]{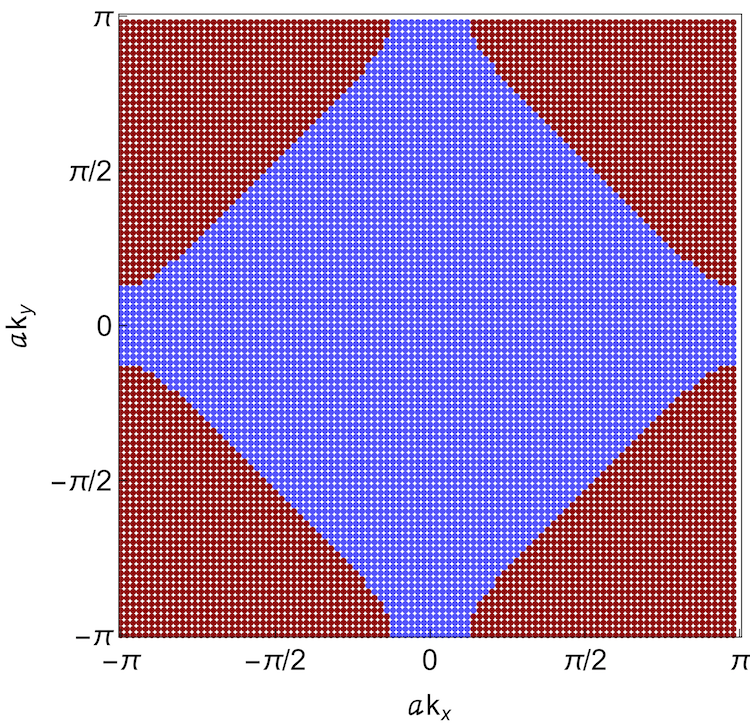}\\

(c)\includegraphics[width=0.4\textwidth]{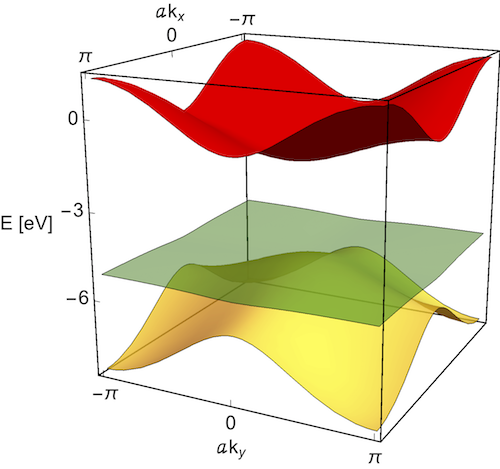}$\qquad$(d)\includegraphics[width=0.4\textwidth]{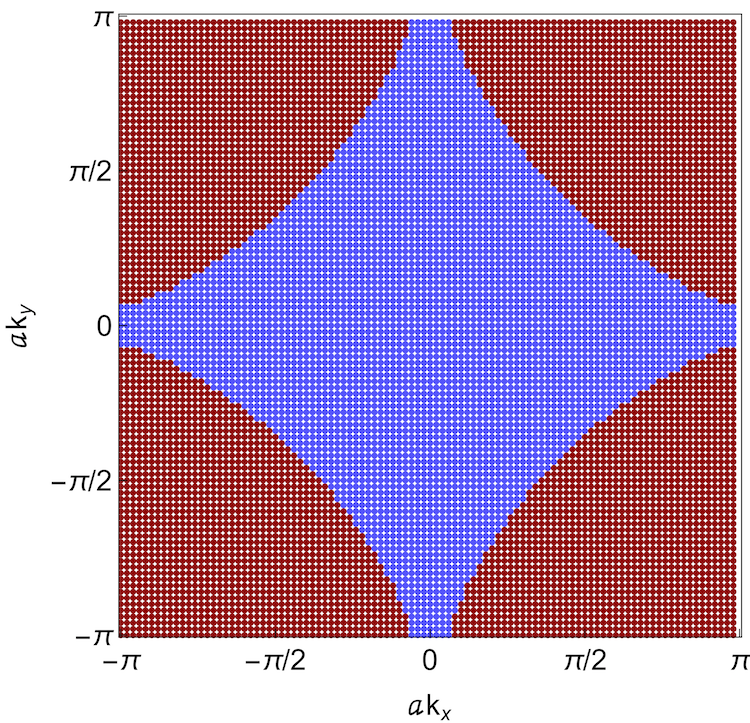}

\caption{Electronic band structure and Fermi surface of a cuprate lattice in
the first Brillouin zone for $V_{dp}=3.45\text{ eV}$ and $t_{pd}=1.13\text{ eV}$:
(a) Band structure for $t_{pp}=0$; (b) the corresponding Fermi surface
for $\mu\approx-0.679\text{ eV}$; (c) band structure for $t_{pp}=0.8\text{ eV}$;
and (d) the corresponding Fermi surface for $\mu\approx-0.679\text{ eV}$
in the upper band. The two lower bands are filled. For $t_{pp}\protect\neq0$
the flat band is deformed, and the curvature of the dispersive bands
is changed. Blue and red dots in panels (b) and (d) correspond to
the occupied and unoccupied states, respectively.}

\label{Fig2}
\end{figure*}

A cuprate lattice is a modified two-dimensional Lieb lattice that
is characterized by a square unit cell with three sites \cite{Lieb_paper,Lieb_lattice_Nita,Plakida_HTc_Book,Leggett_Cuprates};
see Fig. \ref{Fig1}. In terms of single particle terms, we include
potential energies for the \emph{p}- and \emph{d}-orbitals, which
have different values. The difference of these potential energies
modifies the charge-transfer energy. Furthermore, we include a tunneling
energy between the \emph{p}-orbitals. The single-particle energy spectrum
of this lattice has two dispersive bands and one nearly flat band
in between. A large charge-transfer energy provides a gap between
the dispersive upper band and the other two bands, where the dispersion
of the latter is due to the tunneling between \emph{p}-orbitals, which
is assumed to be smaller than the other energies; see Figs. \ref{Fig2}(a)
and \ref{Fig2}(c). As we describe below, we will also assume the
on-site interaction of two particles to be different on \emph{p}-
and \emph{d}-orbitals. Realizations and proposals for this and related
lattices have been reported for cold atoms \cite{BEC_book,trapping_atoms,Lieb_optical_1,Lieb_optical_2,Lieb_optical_3,Lieb_atomic_1,Lieb_atomic_2}
and photonic systems \cite{Lieb_photonic_1} as well as for solid-state
systems \cite{Lieb_material_1,Lieb_material_2}. Phenomena that are
associated with flat-band or three-band systems have been reported
in Refs. \cite{Lieb_lattice_Flach,Flach_flat_band_review,FB_1,FB_2,FB_3,FB_4,FB_5,FB_6}. 

The namesake realization of this lattice are the cuprate materials,
in particular the copper-oxide layers, which are modeled as a cuprate
lattice described above. These layers of Cu and O atoms directly realize
the lattice displayed in Fig. \ref{Fig1}, where the Cu atoms are
represented as a $d_{x^{2}-y^{2}}$ orbital, and O atoms as $p_{x}$
and $p_{y}$ orbitals. It is assumed that the $\mathrm{Cu}\mathrm{O}_{2}$
layers are the location of the electron pairs, and therefore contain
the origin of superconductivity \cite{High_Tc_Handbook,Lynn_HTc_Book,Anderson_Cuprate_Book,Uchida_HTc_Book}.
The mechanism of pairing is under debate, with competing proposals
reported in Refs. \cite{HTc_Phonon_1,HTc_Phonon_2,HTc_Phonon_3,HTc_Phonon_4,HTc_Phonon_5,HTc_Magnon_1,High_Tc_pairing_mechanism_1,HTc_Plasmon_1,HTc_Plasmon_2,HTc_Plasmon_3,HTc_Plasmon_4,Alexandrov_Hubbard_repulsive}.
ARPES measurements have provided insight into the Fermi surface geometry
in the relevant regime \cite{Plakida_HTc_Book,Leggett_Cuprates,Alexandrov_Book,ARPES_1,ARPES_2,ARPES_3,ARPES_4,ARPES_5,Starfish},
which can be captured by the three-band model; see see Fig. \ref{Fig2}(b)
and \ref{Fig2}(d). Numerous theoretical studies on determining the
ground state of the interacting cuprate problem have been reported;
see, e.g., Refs. \cite{Spalek_HTc_Cuprate,Spalek_t_J_model,Spalek_t_J_U_model,H_K_model,DMET_1,DMET_2,AFQMC,iPEPS,DMRG,FLEX_DMFT}.

In the study of conventional superconductors, the Cooper problem and
its solution constituted a defining insight that advanced the understanding
of superconductivity in general. In its original formulation, the
Cooper problem assumes an effective electron-electron attraction due
to the dominance of the electron-phonon interaction over the screened
Coulomb repulsion, leading to a Cooper pair with an orbital \emph{s}-wave
symmetry \cite{Leggett_Cuprates,Cooper_paper,Goodstein_1,Abrikosov_book_metals,Tinkham_1,Aschcroft_Mermin_1}.
This insight paved the way for the subsequent development of BCS theory.
Extensions of the Cooper problem to three-body systems have been presented
in Refs. \cite{Sanayei_1,Sanayei_2}.

In this paper we solve the Cooper problem for a cuprate lattice. The
Cooper problem is intrinsically suitable for the weak-coupling limit
of a Fermi system, because a nearly-intact Fermi sea is assumed. Our
proposal aims primarily at ultracold atom systems, specifically Fermi
mixtures in cuprate optical lattices, which directly realizes the
model that we describe. Utilizing the tunability of these systems,
the interaction and density dependence of the pairing energy can be
measured, to be compared to our predictions. In the weak-coupling
limit, we expect these predictions to be quantitatively accurate.
As we describe below, we formulate the Cooper problem for the cuprate
lattice and solve it numerically, where we emphasize that no assumption
about the orbital symmetry of the ground state is made. We find that
the ground has a $d_{x^{2}-y^{2}}$ symmetry, and that the binding
energy is large for a Fermi surface consisting of four Fermi arcs.
Furthermore, we use the parameters for the three-level model of cuprates
that were reported in Ref. \cite{Spalek_HTc_Cuprate}. We emphasize
that these numbers suggest that the system is in the strongly correlated
regime, and that therefore the Cooper problem approach cannot be expected
to provide quantitatively correct predictions. We present this application
of our calculation firstly for the sake of academic completeness,
because all available methods should be applied to an unsolved problem.
Secondly, we point out that the resulting ground state properties
are consistent with the typical findings in cuprate materials, with
a ground state energy of the order of 100 K.

This paper is organized as follows. In Sec. \ref{Sec_II} we calculate
the electronic band structure of the cuprate lattice for $t_{pp}\neq0$.
We define the Fermi sea, and demonstrate the effect of $t_{pp}$ on
the Fermi-surface geometry. In Sec. \ref{Sec_III} we consider the
Cooper problem, and derive an eigenequation describing a Cooper pair
on the submanifold $\mathcal{S}$ of the upper band. In Sec. \ref{Sec_IV}
we calculate the ground-state energy and wave function, and determine
its orbital symmetry. In Sec. \ref{Sec_V} we propose experimental
signatures of the \emph{d}-wave Cooper pairs for a cold-atom system
in a cuprate lattice. Finally, in Sec. \ref{Sec_VI} we present the
concluding remarks.

\section{electronic band structure and fermi-hubbard model\label{Sec_II}}

For the lattice configuration with three sites A, B, and C in the
square unit cell displayed in Fig. \ref{Fig1}, we assume the on-site
potential to be $V_{\text{A}}\equiv V_{d}$ and $V_{\text{B}}=V_{\text{C}}\equiv V_{p}$.
We also define three sets of creation and annihilation operators $\{a_{nm}^{\dagger},a_{nm}\}$,
$\{b_{nm}^{\dagger},b_{nm}\}$, $\{c_{nm}^{\dagger},c_{nm}\}$ corresponding
to the A-, B-, and C-site, respectively, where the indices $n$ and
$m$ refer to the $x$- and $y$ direction in real space. These operators
fulfill the fermionic algebra, and we refer to them as site operators.
The spin index is suppressed.

The tight-binding Hamiltonian in momentum space is

\begin{equation}
\hat{H}_{\mathrm{tb}}=\sum_{\mathbf{k}\in\,1.\mathrm{BZ}}\left(\begin{array}{ccc}
a_{\mathbf{k}}^{\dagger} & b_{\mathbf{k}}^{\dagger} & c_{\mathbf{k}}^{\dagger}\end{array}\right)\mathrm{h}_{\mathrm{tb}}\left(\begin{array}{c}
a_{\mathbf{k}}\\
b_{\mathbf{k}}\\
c_{\mathbf{k}}
\end{array}\right),\label{H_tb_1}
\end{equation}
for all momentum points $\mathbf{k}=(k_{x},k_{y})$ within the first
Brillouin zone (1.BZ), where $k_{x},k_{y}\in[-\pi/a,\pi/a)$ and $a$
denotes the lattice constant. The matrix $\mathrm{h}_{\mathrm{tb}}$
is

\begin{equation}
\mathrm{h}_{\mathrm{tb}}=\left(\begin{array}{ccc}
V_{d} & f(k_{x}) & -g(k_{y})\\
f^{*}(k_{x}) & V_{p} & -\tau f^{*}(k_{x})g(k_{y})\\
-g^{*}(k_{y}) & -\tau f(k_{x})g^{*}(k_{y}) & V_{p}
\end{array}\right),\label{h_tb}
\end{equation}
where $f(k_{x})=t_{pd}(1-e^{-ik_{x}})$, $g(k_{y})=t_{pd}(1-e^{-ik_{y}})$,
and $\tau=t_{pp}/t_{pd}^{2}$. The parameters $t_{pd}$ and $t_{pp}$
show the nearest-neighbor and next-nearest-neighbor hopping, respectively.
The functions $f^{*}$ and $g^{*}$ denote the complex conjugate of
$f$ and $g$, respectively; see Appendix A. 

The characteristic equation of the matrix $\mathrm{h}_{\mathrm{tb}}$
is cubic with three solutions $E_{\mathbf{k}}^{(\mathrm{U})}$, $E_{\mathbf{k}}^{(\mathrm{F})}$,
and $E_{\mathbf{k}}^{(\mathrm{L})}$, that provide the electronic
band structure of the lattice; see Appendix B for analytical solutions.
Here the index U, F, and L stands for the upper-, flat-, and lower
band, respectively. The dispersion $E_{\mathbf{k}}^{(\mathrm{F})}$
is exactly constant for vanishing $t_{pp}$, resulting in a flat band,
but has a small momentum dependence for nonvanishing $t_{pp}$. However,
we use the index F for both cases. Figures \ref{Fig2}(a) and \ref{Fig2}(c)
show the band structure for vanishing and nonvanishing $t_{pp}$,
respectively. For both cases there are two dispersive bands $E_{\mathbf{k}}^{(\mathrm{U})}$
and $E_{\mathbf{k}}^{(\mathrm{L})}$. For $t_{pp}=0$ the dispersion
is given by $E_{\mathbf{k}}^{(\mathrm{F})}=V_{p}$, between $E_{\mathbf{k}}^{(\mathrm{U})}$
and $E_{\mathbf{k}}^{(\mathrm{L})}$. For $t_{pp}\neq0$ the dispersion
is not constant, however, its momentum dependence is small compared
to the other two bands.

\begin{figure}
\includegraphics[width=1\columnwidth]{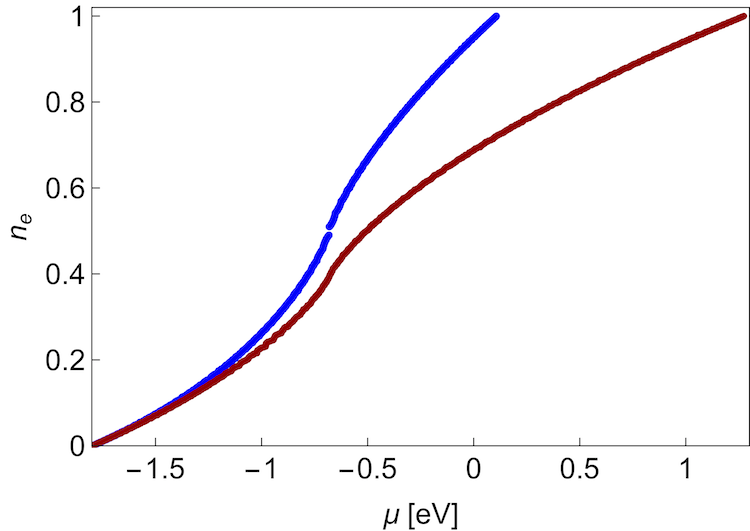}

\caption{Electron density, $n_{e}$, vs chemical potential, $\mu$, in units
of eV, of the upper band of the cuprate lattice, where $V_{dp}=3.45\text{ eV}$
and $t_{pd}=1.13\text{ eV}$. The blue curve corresponds to $t_{pp}=0$,
and the red curve corresponds to $t_{pp}=0.8\text{ eV}$. For a given
value of $\mu$, we have $n_{e}(t_{pp}\protect\neq0)\leqslant n_{e}(t_{pp}=0)$.}

\label{Fig3}
\end{figure}

To formulate the Cooper problem for the upper band, we consider two-particle
states with vanishing total momentum. To find the interaction term
of the Cooper problem, we use the three sets of creation and annihilation
operators $\{\psi_{\mathrm{U},\mathbf{k}\sigma}^{\dagger},\psi_{\mathrm{U},\mathbf{k}\sigma}\}$,
$\{\psi_{\mathrm{F},\mathbf{k}\sigma}^{\dagger},\psi_{\mathrm{F},\mathbf{k}\sigma}\}$,
and $\{\psi_{\mathrm{L},\mathbf{k}\sigma}^{\dagger},\psi_{\mathrm{L},\mathbf{k}\sigma}\}$,
where $\sigma\in\{\uparrow,\downarrow\}$ is a spin index. These operators
fulfill the fermionic algebra, and create or annihilate an electron
in the the upper-, flat-, and lower band, respectively. In the following
we refer to them as band operators. The band operators can be related
to the site operators using the components of the eigenvectors of
the matrix $\mathrm{h}_{\mathrm{tb}}$; see Appendix C. We assume
that the charge-transfer energy, $V_{dp}\equiv V_{d}-V_{p}$, is sufficiently
large so that we neglect the interband pairings; cf. Figs. \ref{Fig2}(a)
and \ref{Fig2}(c). We write the interaction Hamiltonian for the Cooper
problem as

\begin{align}
\hat{H}_{\mathrm{int}}= & \frac{1}{\mathcal{A}}\sum_{\mathbf{k},\mathbf{k}'\in\text{1.BZ}}\mathcal{V}_{\mathbf{k},\mathbf{k}'}\psi_{\mathrm{U},\mathbf{k}'\downarrow}^{\dagger}\psi_{\mathrm{U},-\mathbf{k}'\uparrow}^{\dagger}\psi_{\mathrm{U},-\mathbf{k}\uparrow}\psi_{\mathrm{U},\mathbf{k}\downarrow};\label{H_int}
\end{align}
see Appendix C for the derivation. Here, $\mathcal{A}$ denotes the
area of the first Brillouin zone and the interaction function $\mathcal{V}_{\mathbf{k},\mathbf{k}'}$
is
\begin{equation}
\mathcal{V}_{\mathbf{k},\mathbf{k}'}=U_{d}\mathcal{V}_{\mathbf{k},\mathbf{k}'}^{(d)}+U_{p}\left(\mathcal{V}_{\mathbf{k},\mathbf{k}'}^{(p_{x})}+\mathcal{V}_{\mathbf{k},\mathbf{k}'}^{(p_{y})}\right),\label{V_k_kprime}
\end{equation}
where the functions $\mathcal{V}_{\mathbf{k},\mathbf{k}'}^{(d)}$,
$\mathcal{V}_{\mathbf{k},\mathbf{k}'}^{(p_{x})}$, and $\mathcal{V}_{\mathbf{k},\mathbf{k}'}^{(p_{y})}$
are derived in Appendix C. The on-site Coulomb interaction strength
for $d_{x^{2}-y^{2}}$ orbitals is $U_{d}$, and for both $p_{x}$
and $p_{y}$ orbitals it is $U_{p}$. 

\begin{figure}
\includegraphics[width=0.95\columnwidth]{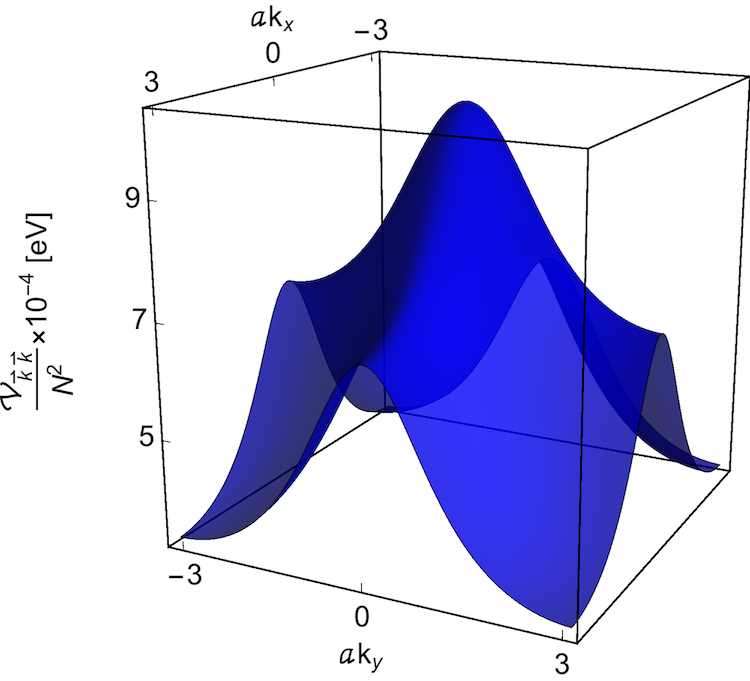}

\caption{Interaction function $\mathcal{V}_{\mathbf{k},\mathbf{k}}/N^{2}$
in units of eV, cf. Eq. (\ref{V_k_kprime}), for repulsive on-site
interactions, where $N=100$, $V_{dp}=3.45\text{ eV}$, $t_{pd}=1.13\text{ eV}$,
$t_{pp}=0.8\text{ eV}$, $U_{d}=10.3\text{ eV}$, and $U_{p}=4.1\text{ eV}.$}

\label{Fig4}
\end{figure}

Next, we define the Fermi sea by introducing a chemical potential,
$\mu$. We define the interacting Fermi sea, $\mathrm{FS_{\mathrm{int}}}$,
which includes corrections due to the interaction:

\begin{equation}
\mathrm{FS_{\mathrm{int}}=}\Bigl\{\mathbf{k}\in\text{1.BZ}:2E_{\mathbf{k}}^{(\mathrm{U})}+\frac{1}{\mathcal{A}}\mathcal{V}_{\mathbf{k},\mathbf{k}}<2\mu\Bigr\},\label{FS_interacing}
\end{equation}
where the interaction function $\mathcal{V}_{\mathbf{k},\mathbf{k}}$
is obtained as Eq. (\ref{V_k_kprime}) for $\mathbf{k}'=\mathbf{k}$;
see Fig. \ref{Fig4}. We assume these states to be occupied with an
inert Fermi sea. The unoccupied states that are considered in the
Cooper problem are the Brillouin zone without the Fermi sea; i.e.,
$\mathbf{k}\in\text{\text{1.BZ}\ensuremath{\setminus\mathrm{FS_{\mathrm{int}}}}}$.
Figures \ref{Fig2}(b) and \ref{Fig2}(d) show the Fermi surface for
$t_{pp}=0$ and $t_{pp}\neq0$, respectively. The nonvanishing $t_{pp}$
changes the curvature of the dispersive bands, resulting in a Fermi-surface
geometry that is in better agreement with the experimental data extracted
from ARPES \cite{Plakida_HTc_Book,Leggett_Cuprates,Alexandrov_Book,ARPES_1,ARPES_2,ARPES_3,ARPES_4,ARPES_5,Starfish}.
Moreover, we vary $\mu$ and calculate the corresponding electron
density, $n_{e}$, for both $t_{pp}=0$ and $t_{pp}\neq0$, resulting
in a chemical potential dependence of the density shown in Fig. \ref{Fig3}.
For a given value of $\mu$ we find that $n_{e}(t_{pp}\neq0)\leqslant n_{e}(t_{pp}=0)$.
As a result, while the desired geometry of the Fermi surface is preserved,
we can increase the hole doping for $t_{pp}\neq0$ \cite{hole_doping}. 

Finally, we include the kinetic energy, so that the total Hamiltonian
of the Cooper problem is

\begin{align}
\hat{H}_{\mathrm{tot}}= & \mathop{\sum_{\mathbf{k}\in\,\text{1.BZ}\setminus\mathrm{FS_{\mathrm{int}}}}}_{\sigma\in\{\uparrow,\downarrow\}}\xi_{\mathbf{k}}^{(\mathrm{U})}\psi_{\mathrm{U},\mathbf{k}\sigma}^{\dagger}\psi_{\mathrm{U},\mathbf{k}\sigma}+\frac{1}{\mathcal{A}}\nonumber \\
 & \times\sum_{\mathbf{k},\mathbf{k}'\in\,\text{1.BZ}\setminus\mathrm{FS_{\mathrm{int}}}}\mathcal{V}_{\mathbf{k},\mathbf{k}'}\psi_{\mathrm{U},\mathbf{k}'\downarrow}^{\dagger}\psi_{\mathrm{U},-\mathbf{k}'\uparrow}^{\dagger}\psi_{\mathrm{U},-\mathbf{k}\uparrow}\psi_{\mathrm{U},\mathbf{k}\downarrow},\label{H_tot}
\end{align}
where $\xi_{\mathbf{k}}^{(\mathrm{U})}=E_{\mathbf{k}}^{(\mathrm{U})}-\mu$
and $\mathcal{V}_{\mathbf{k},\mathbf{k}'}$ is given by Eq. (\ref{V_k_kprime});
see Appendix C. 

\section{cooper problem and pairing equation\label{Sec_III}}

The original Cooper problem and its solution show that two electrons
that are immersed in an inert Fermi sea form a bound state with an
orbital \emph{s}-wave symmetry for an arbitrarily weak attractive
interaction \cite{Leggett_Cuprates,Cooper_paper,Goodstein_1,Abrikosov_book_metals,Tinkham_1,Aschcroft_Mermin_1}.
The effective attraction models the phonon-mediated interaction that
is dominant over the screened Coulomb repulsion. The interaction model
that is used for conventional superconductors is a negative coupling
constant in momentum space for the relative kinetic energy of the
electrons smaller than the Debye energy; cf. e.g., Refs. \cite{Cooper_paper,Tinkham_1,Aschcroft_Mermin_1}.
Following the standard Cooper problem, we consider a singlet-state
Cooper pair as

\begin{equation}
\left|\Phi\right\rangle =\sum_{\boldsymbol{\kappa}\in\text{1.BZ}\setminus\mathrm{FS_{int}}}\phi(\boldsymbol{\kappa})\psi_{\mathrm{U},\boldsymbol{\kappa}\uparrow}^{\dagger}\psi_{\mathrm{U},-\boldsymbol{\kappa}\downarrow}^{\dagger}\left|\mathrm{FS_{int}}\right\rangle ,\label{Cooper_pairing_ansatz}
\end{equation}
where $\phi(\boldsymbol{\kappa})$ is the wave function of the Cooper
pair in momentum space and $\left|\mathrm{FS_{int}}\right\rangle $
denotes the Fermi-sea state; cf. Eq. (\ref{FS_interacing}).

\begin{figure}
\includegraphics[width=1\columnwidth]{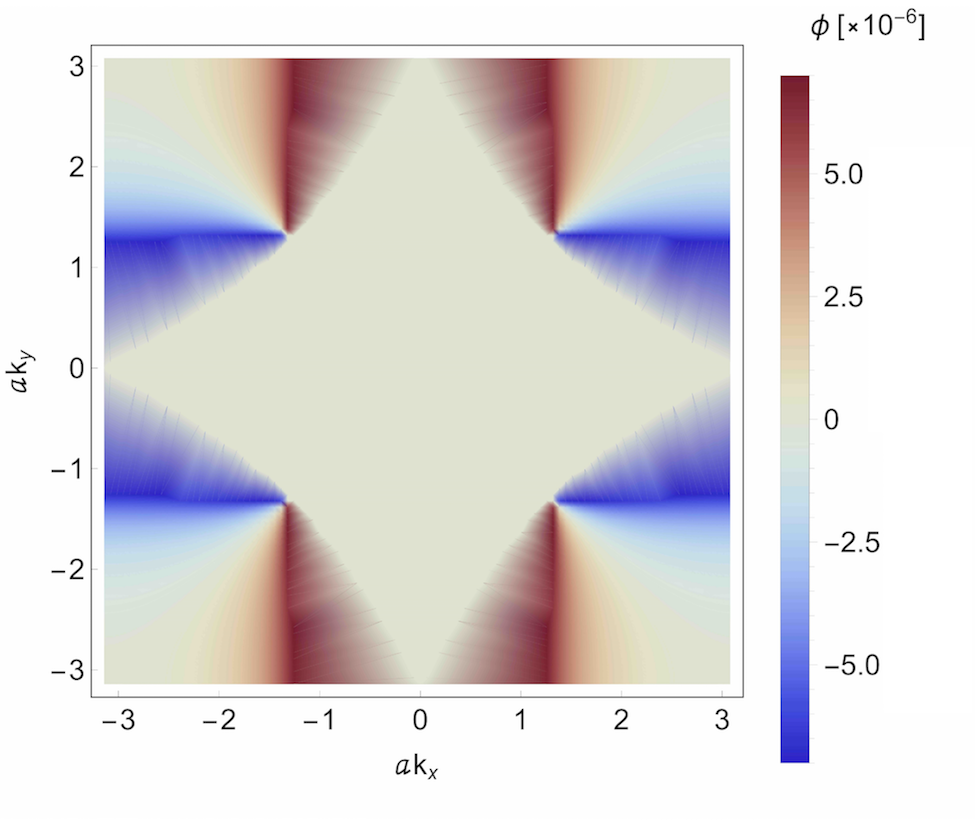}

\caption{Ground-state wave function, $\phi(ak_{x},ak_{y})$, of a Cooper pair
for $N=100$, $V_{dp}=3.45\text{ eV}$, $t_{pd}=1.13\text{ eV}$,
$t_{pp}=0.8\text{ eV}$, $U_{d}=10.3\text{ eV}$, $U_{p}=4.1\text{ eV}$,
and the hole doing $\delta_{h}\approx0.104$, where $a$ denotes the
lattice constant. The nodal points are visible along the Fermi arcs.
The blue color indicates positive values of the wave function, the
red color negative ones. The orbital symmetry of the wave function
is $d_{x^{2}-y^{2}}$.}

\label{Fig5}
\end{figure}

To find the ground-state energy and wave function, we consider the
eigenvalue problem

\begin{equation}
\hat{H}_{\mathrm{tot}}\left|\Phi\right\rangle =\mathcal{E}\left|\Phi\right\rangle ,\label{eigen_Eq_1}
\end{equation}
where $\mathcal{E}$ is the eigenenergy. We determine the operator
$\hat{H}_{\mathrm{tot}}\left|\Phi\right\rangle $ and obtain an eigenequation
describing the Cooper pair:

\begin{align}
\left(\xi_{\mathbf{k}}^{(\mathrm{U})}+\xi_{-\mathbf{k}}^{(\mathrm{U})}+\frac{1}{\mathcal{A}}\mathcal{V}_{\mathbf{k},\mathbf{k}}-\mathcal{E}\right)\phi(\mathbf{k})\qquad\qquad\qquad\qquad\nonumber \\
=-\frac{1}{\mathcal{A}}\mathop{\sum_{\mathbf{k},\mathbf{k}'\in\mathrm{1.BZ}\setminus\mathrm{FS_{int}}}}_{\mathbf{k}'\neq\mathbf{k}}\mathcal{V}_{\mathbf{k},\mathbf{k}'}\phi(\mathbf{k}');\label{eigen_Eq_2}
\end{align}
see Appendix D. In what fallows we solve Eq. (\ref{eigen_Eq_2}) numerically,
and determine the ground-state energy, $E_{\text{G}}<0$.

\section{ground-state energy and wave function\label{Sec_IV}}

To solve Eq. (\ref{eigen_Eq_2}) numerically, first we discretize
the first Brillouin zone as $\mathbf{k}_{j}=(k_{x}^{(j)},k_{y}^{(j)})$,
where

\begin{equation}
k_{x}^{(j)},k_{y}^{(j)}=\frac{1}{a}\bigl[-\pi+\frac{2\pi}{N}(j-1)\bigr]\text{ }\text{ for }j=1,2,\ldots,N.\label{grid_points}
\end{equation}
Here, $a$ denotes the lattice constant and $N\in\mathbb{N}$ is the
number of grid points in \emph{$x$}- and \emph{$y$} direction, i.e.,
$N_{x}=N_{y}=N$. We calculate the electronic band structure numerically
and the interaction function $\mathcal{V}_{\mathbf{k},\mathbf{k}'}$
at each grid point using Eq. (\ref{V_k_kprime}). Next, for a given
value of $\mu$ we determine the Fermi surface following the relation
(\ref{FS_interacing}). We note that the number of grid points within
the first Brillouin zone is proportional to $N_{x}N_{y}=N^{2}$, so
the size of the matrix associated with $\hat{H}_{\mathrm{tot}}$ is
proportional to $N^{4}$, cf. Eq. (\ref{H_tot}). We choose the number
of grid points sufficiently large to ensure convergence of the numerical
results. We determine the Fermi sea numerically, and exclude it from
the first Brillouin zone. We consider $\hat{H}_{\mathrm{tot}}$ on
the reduced momentum space that corresponds to the unoccupied states;
see Appendix E.

For the regime of attractive interactions, i.e., $U_{d},U_{p}<0$,
we find a Cooper pair with an approximate orbital \emph{s}-wave symmetry;
see Appendix F. 

For repulsive on-site interactions, $U_{d},U_{p}>0$, the interaction
function $\mathcal{V}_{\mathbf{k},\mathbf{k}}$, cf. Eq. (\ref{V_k_kprime}),
has the momentum dependence shown in Fig. \ref{Fig4}. The repulsive
interaction suppresses the formation of pairs with \emph{s}-wave symmetry.
Instead, the ground state wave function has d-wave symmetry, as shown
in Fig. \ref{Fig5}, for the lattice parameters $V_{dp}=3.45\text{ eV}$,
$t_{pd}=1.13\text{ eV}$, and $t_{pp}=0.8\text{ eV}$ that follow
approximately the values given by Ref. \cite{Spalek_HTc_Cuprate}.
The orientation of the maxima and minima, and the location of the
nodal points, indicates that the wave function supports an orbital
symmetry of $d_{x^{2}-y^{2}}$.

\begin{figure}[t]
\includegraphics[width=1\columnwidth]{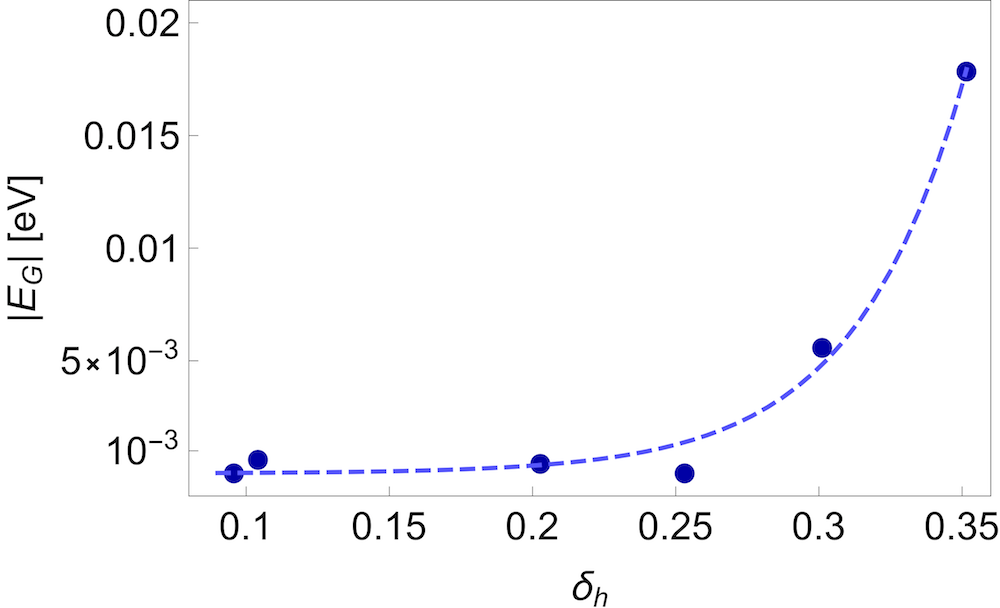}

\caption{Ground-state energy $|E_{\text{G}}|$ of a Cooper pair in units of
eV vs hole doping, $\delta_{h}$ \cite{hole_doping}, where $N=100$,
$V_{dp}=3.45\text{ eV}$, $t_{pd}=1.13\text{ eV}$, $t_{pp}=0.8\text{ eV}$,
$U_{d}=10.3\text{ eV}$, and $U_{p}=4.1\text{ eV}$. A largest value
of $|E_{\text{G}}|$ corresponding to a critical temperature of the
order of 100 K is obtained near the hole doping of 0.35. }

\label{Fig6}
\end{figure}

Finally, we vary the hole doping, $\delta_{h}$, by changing the chemical
potential, $\mu$ \cite{hole_doping}. We calculate the corresponding
ground-state energy, $E_{\text{G}}$, resulting in Fig. \ref{Fig6}.
We find that the largest absolute magnitude of the ground-state energy,
$|E_{\text{G}}^{(\text{max})}|\sim0.01\text{ eV}$, occurs near the
hole doping $\delta_{h}\sim0.35$. This converts to a critical temperature
of the order of $100\text{ K}$. The behavior of the ground-state
energies consistent with Ref. \cite{Spalek_HTc_Cuprate} qualitatively. 

\section{experimental signature in a cold-atom system\label{Sec_V}}

We propose to detect the predictions of our analysis in a system of
ultracold atoms in an optical lattice. Specifically, we consider fermionic
atoms in higher bands of optical lattices. Lattice geometries that
resemble the cuprate lattice and related geometries have been realized
experimentally for bosonic atoms in Refs. \cite{Lieb_optical_1,Lieb_optical_2,Lieb_optical_3}.
Utilizing a Feshbach resonance, the whole range of repulsive interactions
is accessible, from weak to strong coupling. In particular, our predictions
can be tested quantitatively in the weak-coupling regime. 

We propose to use noise correlations of time-of-flight images as an
observable to detect the symmetry of the Cooper pair. In the far-field
limit, which is achieved for expansions in which the expanded cloud
is much larger than the in-situ cloud, the density correlations of
the atoms of different spins include the correlations of $\hat{n}(\mathbf{k},\sigma)$
and $\hat{n}(-\mathbf{k},-\sigma)$, where $\hat{n}(\mathbf{k},\sigma)$
is the occupation of the momentum state $\mathbf{k}$ and spin-state
$\sigma$ of the in-situ system; see Refs. \cite{time_of_flight_1,time_of_flight_2,noise_correlation_1,noise_correlation_2}.
This quantity gives access to the square of the pair wave function,
depicted in Fig. \ref{Fig5}. In particular the angular dependence
of its magnitude and the nodal points of the wave function are observable
in this quantity.

As a second measurement, we propose to use stirring experiments, as
discussed in Ref. \cite{noise_correlation_3}. Here, either a focused
laser beam is moved through the quantum gas, or a lattice is dragged
through it. For fermionic systems, the heating that is induced by
this perturbation is suppressed for stirring velocities smaller than
the critical velocity $\sim\Delta_{\mathbf{k}_{F}}/|\mathbf{k}_{F}|$.
Here, the energy gap $\Delta_{\mathbf{k}_{F}}$ refers to the gap
at a momentum $\mathbf{k}_{F}$ that is on the Fermi surface and along
the direction of the motion of the stirring potential. Therefore the
heating rate and its dependence on the stirring direction maps out
the energy gap of the paired state.

\section{conclusions\label{Sec_VI}}

In conclusion, we have presented the solution of the Cooper problem
for a cuprate lattice, for repulsive interactions. The band structure
of the cuprate lattice consists of three bands, where we focus on
densities for which the two lower bands are filled, and the Fermi
surface is in the highest band. For these densities, and for repulsive
on-site interactions, we demonstrate that the ground state solution
of the Cooper problem has a $d_{x^{2}-y^{2}}$ orbital symmetry. The
binding energy of the Cooper pair depends strongly on the shape of
the Fermi surface. We show that it is small for a connected surface
of the shape of a deformed circle, while it is large for surfaces
that break up into four disconnected arcs. As a primary, quantitative
implementation of our results we propose to create an ultracold Fermi
gas in a cuprate lattice. Here, the weak-coupling regime can be implemented
naturally due to the tunable nature of these systems, and the dependence
on system parameters such as the interaction strength and the Fermi
surface geometry mapped out. We pointed out noise correlations and
stirring experiments as experimental methods to detect our predictions.
As a second platform, we apply our calculation to the three-band model
reported for cuprate materials. We emphasize that the interaction
strengths of this model suggests that it is in the strongly correlated
regime, whereas the Cooper problem is primarily applicable in the
weak-coupling limit. However, we present the predictions of the Cooper
problem here, given the impact of the Cooper problem on the study
of superconductivity, primarily for the purpose of academic completeness.
We find that the solution of Cooper problem predicts the experimentally
observed $d_{x^{2}-y^{2}}$ symmetry of the electron pairs, a sharp
increase of the binding energy with increasing hole doping when the
Fermi surface breaks up into four disconnected arcs, and a maximal
binding energy of approximately 100 K. This study and its experimental
implementation provides a direct analogy between cold-atom systems
and the three-band model that is utilized in cuprate materials, and
therefore advances the exchange between cold-atom and condensed-matter
systems. 

\subsection*{ACKNOWLEDGMENTS }

This work was funded by the Deutsche Forschungsgemeinschaft (DFG,
German Research Foundation) \textendash{} SFB-925 \textendash{} project
170620586, and by the Cluster of Excellence \textquoteleft Advanced
Imaging of Matter\textquoteright{} of the Deutsche Forschungsgemeinschaft
(DFG) EXC 2056 - project ID 390715994.

\setcounter{equation}{0} \renewcommand{\theequation}{A\arabic{equation}}

\section*{appendix a. derivation of the tight-binding hamiltonian (\ref{H_tb_1})}

We consider the cuprate lattice, see Fig. \ref{Fig1}, and write the
spinless tight-binding Hamiltonian in terms of the site operators
in real space:

\begin{align}
\hat{H}_{\mathrm{tb}}= & \sum_{nm}\Bigl[V_{d}a_{nm}^{\dagger}a_{nm}+V_{p}b_{nm}^{\dagger}b_{nm}+V_{d}c_{nm}^{\dagger}c_{nm}\nonumber \\
 & \quad+t_{pd}a_{nm}^{\dagger}b_{nm}+t_{pd}b_{nm}^{\dagger}a_{nm}-t_{pd}a_{nm}^{\dagger}c_{nm}\nonumber \\
 & \quad-t_{pd}c_{nm}^{\dagger}a_{nm}-t_{pd}a_{nm}^{\dagger}b_{n-1,m}-t_{pd}b_{n-1,m}^{\dagger}a_{nm}\nonumber \\
 & \quad+t_{pd}a_{nm}^{\dagger}c_{n,m-1}+t_{pd}c_{n,m-1}^{\dagger}a_{nm}-t_{pp}b_{nm}^{\dagger}c_{nm}\nonumber \\
 & \quad-t_{pp}c_{nm}^{\dagger}b_{nm}+t_{pp}c_{nm}^{\dagger}b_{n-1,m}+t_{pp}b_{n-1,m}^{\dagger}c_{nm}\nonumber \\
 & \quad-t_{pp}b_{n-1,m}^{\dagger}c_{n,m-1}-t_{pp}c_{n,m-1}^{\dagger}b_{n-1,m}\nonumber \\
 & \quad+t_{pp}b_{n,m}^{\dagger}c_{n,m-1}+t_{pp}c_{n,m-1}^{\dagger}b_{n,m}\Bigr],\label{H_tb_derivation_1}
\end{align}
where $n$ and $m$ are two indices for the $x$- and $y$ direction,
respectively. Next, we take the Fourier transform of each operator,
and obtain the tight-binding Hamiltonian in momentum space:

\begin{align}
\hat{H}_{\mathrm{tb}}= & \sum_{\mathbf{k}\in\,1.\mathrm{BZ}}\Bigl[V_{d}a_{\mathbf{k}}^{\dagger}a_{\mathbf{k}}+V_{p}b_{\mathbf{k}}^{\dagger}b_{\mathbf{k}}+V_{d}c_{\mathbf{k}}^{\dagger}c_{\mathbf{k}}\nonumber \\
 & \quad\quad+t_{pd}a_{\mathbf{k}}^{\dagger}b_{\mathbf{k}}+t_{pd}b_{\mathbf{k}}^{\dagger}a_{\mathbf{k}}-t_{pd}a_{\mathbf{k}}^{\dagger}c_{\mathbf{k}}\nonumber \\
 & \quad\quad-t_{pd}c_{\mathbf{k}}^{\dagger}a_{\mathbf{k}}-t_{pd}e^{-ik_{x}}a_{\mathbf{k}}^{\dagger}b_{\mathbf{k}}-t_{pd}e^{ik_{x}}b_{\mathbf{k}}^{\dagger}a_{\mathbf{k}}\nonumber \\
 & \quad\quad+t_{pd}e^{-ik_{y}}a_{\mathbf{k}}^{\dagger}c_{\mathbf{k}}+t_{pd}e^{ik_{y}}c_{\mathbf{k}}^{\dagger}a_{\mathbf{k}}\nonumber \\
 & \quad\quad-t_{pp}\left(1-e^{ik_{x}}+e^{ik_{x}}e^{-ik_{y}}-e^{-ik_{y}}\right)b_{\mathbf{k}}^{\dagger}c_{\mathbf{k}}\nonumber \\
 & \quad\quad-t_{pp}\left(1-e^{-ik_{x}}+e^{-ik_{x}}e^{ik_{y}}-e^{ik_{y}}\right)c_{\mathbf{k}}^{\dagger}b_{\mathbf{k}}\Bigr].\label{H_tb_derivation_2}
\end{align}
Finally, we define $f(k_{x})=t_{pd}(1-e^{-ik_{x}})$, $g(k_{y})=t_{pd}(1-e^{-ik_{y}})$,
and $\tau=t_{pp}/t_{pd}^{2}$, and arrive at Eq. (\ref{H_tb_1}).

\setcounter{equation}{0} \renewcommand{\theequation}{B\arabic{equation}}

\section*{appendix b. analytical description of the band structure of the cuprate
lattice}

The characteristic equation associated with Eq. (\ref{h_tb}) reads
as

\begin{equation}
s^{3}+c(k_{x},k_{y})s^{2}+d(k_{x},k_{y})s+e(k_{x},k_{y})=0,\label{charac_Eq_1}
\end{equation}
where

\begin{align}
c(k_{x},k_{y})= & -V_{d}-2V_{p},\label{c_func}
\end{align}

\begin{align}
d(k_{x},k_{y})= & -|f(k_{x})|^{2}-|g(k_{y})|^{2}-\tau^{2}|f(k_{x})|^{2}|g(k_{y})|^{2}\nonumber \\
 & +V_{p}^{2}+2V_{d}V_{p},\label{d_func}
\end{align}

\begin{align}
e(k_{x},k_{y})= & V_{p}|f(k_{x})|^{2}+V_{p}|g(k_{y})|^{2}+\tau(\tau V_{d}-2)\nonumber \\
 & \times|f(k_{x})|^{2}|g(k_{y})|^{2}-V_{d}V_{p}^{2},\label{e_func}
\end{align}
and $\tau=t_{pp}/t_{pd}^{2}$. Next, we define a variable $S=s-c(k_{x},k_{y})/3$,
and rewrite Eq. (\ref{charac_Eq_1}) as

\begin{equation}
S^{3}+3p(k_{x},k_{y})S+2q(k_{x},k_{y})=0,\label{charac_Eq_2}
\end{equation}
where

\begin{align}
p(k_{x},k_{y})= & \frac{1}{3}d(k_{x},k_{y})-\frac{1}{9}[c(k_{x},k_{y})]^{2},\label{p_func}
\end{align}

\begin{align}
q(k_{x},k_{y})= & \frac{1}{27}[c(k_{x},k_{y})]^{3}-\frac{1}{6}c(k_{x},k_{y})d(k_{x},k_{y})\nonumber \\
 & +\frac{1}{2}e(k_{x},k_{y}).\label{q_func}
\end{align}
Following the mathematical formalism represented in Ref. \cite{cubic_eq},
we calculate the three roots of Eq. (\ref{charac_Eq_1}), revealing
the band structure of the cuprate lattice:

\begin{align}
E_{\mathbf{k}}^{(\mathrm{U})}= & 2\sqrt{-p(k_{x},k_{y})}\cos\left(\frac{\theta(k_{x},k_{y})}{3}\right)-\frac{c(k_{x},k_{y})}{3},\label{E_U_full}
\end{align}

\begin{align}
E_{\mathbf{k}}^{(\mathrm{F})}= & 2\sqrt{-p(k_{x},k_{y})}\cos\left(\frac{\theta(k_{x},k_{y})+4\pi}{3}\right)-\frac{c(k_{x},k_{y})}{3},\label{E_F_full}
\end{align}

\begin{align}
E_{\mathbf{k}}^{(\mathrm{L})}= & 2\sqrt{-p(k_{x},k_{y})}\cos\left(\frac{\theta(k_{x},k_{y})+2\pi}{3}\right)-\frac{c(k_{x},k_{y})}{3},\label{E_L_full}
\end{align}
where $\cos\theta(k_{x},k_{y})=-q(k_{x},k_{y})/\sqrt{-[p(k_{x},k_{y})]^{3}}$;
see Fig. \ref{Fig2}(c). We note that for $t_{pp}=0$, the three solutions
(\ref{E_U_full})-(\ref{E_L_full}) reduce to:

\begin{alignat}{1}
\tilde{E}_{\mathbf{k}}^{(\mathrm{U})}= & \frac{V_{d}+V_{p}}{2}+2t_{pd}\sqrt{\sin^{2}(\frac{k_{x}}{2})+\sin^{2}(\frac{k_{y}}{2})+\left(\frac{V_{dp}}{4t_{pd}}\right)^{2}},\label{E_U_tpp_zero}
\end{alignat}

\begin{alignat}{1}
\tilde{E}_{\mathbf{k}}^{(\mathrm{F})}= & V_{p},\label{E_F_tpp_zero}
\end{alignat}

\begin{alignat}{1}
\tilde{E}_{\mathbf{k}}^{(\mathrm{L})}= & \frac{V_{d}+V_{p}}{2}-2t_{pd}\sqrt{\sin^{2}(\frac{k_{x}}{2})+\sin^{2}(\frac{k_{y}}{2})+\left(\frac{V_{dp}}{4t_{pd}}\right)^{2}},\label{E_L_tpp_zero}
\end{alignat}
respectively; see Fig. \ref{Fig2}(a). By comparing Eqs. (\ref{E_U_full})-(\ref{E_L_tpp_zero})
we find that the next-nearest-neighbor hopping, $t_{pp}$, deforms
the flat band $E_{\mathbf{k}}^{(\mathrm{F})}$, and changes the curvature
of the dispersive bands $E_{\mathbf{k}}^{(\mathrm{U})}$ and $E_{\mathbf{k}}^{(\mathrm{L})}$. 

\setcounter{equation}{0} \renewcommand{\theequation}{C\arabic{equation}}

\section*{appendix c. derivation of the hamiltonians (\ref{H_int}) and (\ref{H_tot})}

For the cuprate lattice, see Fig. \ref{Fig1}, the interaction Hamiltonian
of the Fermi-Hubbard model reads in general as

\begin{alignat}{1}
\hat{\tilde{H}}_{\mathrm{int}}= & \frac{U_{\mathrm{C}}}{\mathcal{A}}\sum_{\mathbf{k},\mathbf{p},\mathbf{q}\in\,\text{1.BZ}}\alpha_{\mathbf{k}+\mathbf{q},\downarrow}^{\dagger}\alpha_{\mathbf{p}-\mathbf{q},\uparrow}^{\dagger}\alpha_{\mathbf{p},\uparrow}\alpha_{\mathbf{k}\downarrow},\label{H_int_general}
\end{alignat}
where $\alpha^{\dagger}\in\{a^{\dagger},b^{\dagger},c^{\dagger}\}$
and $\alpha\in\{a,b,c\}$ denote the creation and annihilation site
operators, respectively, $\mathbf{q}$ is the momentum transfer \cite{momentum_transfer},
and $U_{\mathrm{C}}$ is an on-site Coulomb interaction strength.
We notice that for each eigenvalue of $\mathrm{h}_{\mathrm{tb}}$,
cf. Eq. (\ref{h_tb}), there exists a corresponding normalized eigenvector,
which we denote as $\mathbf{v}_{\mathbf{k}}^{(\mathrm{U})}=(v_{\mathbf{k}}^{(1;\mathrm{U})},v_{\mathbf{k}}^{(2;\mathrm{U})},v_{\mathbf{k}}^{(3;\mathrm{U})})$,
$\mathbf{v}_{\mathbf{k}}^{(\mathrm{F})}=(v_{\mathbf{k}}^{(1;\mathrm{F})},v_{\mathbf{k}}^{(2;\mathrm{F})},v_{\mathbf{k}}^{(3;\mathrm{F})})$,
and $\mathbf{v}_{\mathbf{k}}^{(\mathrm{L})}=(v_{\mathbf{k}}^{(1;\mathrm{L})},v_{\mathbf{k}}^{(2;\mathrm{L})},v_{\mathbf{k}}^{(3;\mathrm{L})})$.
The index U, F, and L corresponds to the upper-, flat-, and lower
band, respectively. The site operators can be related to the band
operators using the following relation:

\begin{alignat}{1}
\left(\begin{array}{c}
a_{\mathbf{\mathbf{k}\sigma}}^{\dagger}\\
b_{\mathbf{k}\sigma}^{\dagger}\\
c_{\mathbf{k}\sigma}^{\dagger}
\end{array}\right)= & \left(\begin{array}{ccc}
v_{\mathbf{k}}^{(1;\mathrm{U})} & v_{\mathbf{k}}^{(2;\mathrm{U})} & v_{\mathbf{k}}^{(3;\mathrm{U})}\\
v_{\mathbf{k}}^{(1;\mathrm{F})} & v_{\mathbf{k}}^{(2;\mathrm{F})} & v_{\mathbf{k}}^{(3;\mathrm{F})}\\
v_{\mathbf{k}}^{(1;\mathrm{L})} & v_{\mathbf{k}}^{(2;\mathrm{L})} & v_{\mathbf{k}}^{(3;\mathrm{L})}
\end{array}\right)^{-1}\left(\begin{array}{c}
\psi_{\mathrm{U},\mathbf{k}\sigma}^{\dagger}\\
\psi_{\mathrm{F},\mathbf{k}\sigma}^{\dagger}\\
\psi_{\mathrm{L},\mathbf{k}\sigma}^{\dagger}
\end{array}\right)\nonumber \\
\equiv & \left(\begin{array}{ccc}
v_{11}(\mathbf{k}) & v_{12}(\mathbf{k}) & v_{13}(\mathbf{k})\\
v_{21}(\mathbf{k}) & v_{22}(\mathbf{k}) & v_{23}(\mathbf{k})\\
v_{31}(\mathbf{k}) & v_{32}(\mathbf{k}) & v_{33}(\mathbf{k})
\end{array}\right)\left(\begin{array}{c}
\psi_{\mathrm{U},\mathbf{k}\sigma}^{\dagger}\\
\psi_{\mathrm{F},\mathbf{k}\sigma}^{\dagger}\\
\psi_{\mathrm{L},\mathbf{k}\sigma}^{\dagger}
\end{array}\right).\label{site_op_interms_band_op}
\end{alignat}
We can rewrite the interaction Hamiltonian (\ref{H_int_general})
corresponding to three sites A, B, and C in terms of the band operators
using the relation (\ref{site_op_interms_band_op}). We recall that
here we are primarily interested in a submanifold $\mathcal{S}$,
where the total momentum of an electron-pair is vanishing. Because
we are interested in the effective Fermi-Hubbard model constituted
in the upper band, we prevent the interband pairings as well as the
pairings in the flat- and lower band. We write the three interaction
Hamiltonians corresponding to $d_{x^{2}-y^{2}},$ $p_{x}$, and $p_{y}$
orbital configurations on the submanifold $\mathcal{S}$ of the upper
band in terms of the band operators:
\begin{align}
\hat{H}_{\mathrm{int}}^{(\Omega)}= & \frac{U_{\Omega}}{\mathcal{A}}\sum_{\mathbf{k},\mathbf{k}'\in\,\text{1.BZ}}\mathcal{V}_{\mathbf{k},\mathbf{k}'}^{(\Omega)}\psi_{\mathrm{U},\mathbf{k}'\downarrow}^{\dagger}\psi_{\mathrm{U},-\mathbf{k}'\uparrow}^{\dagger}\psi_{\mathrm{U},-\mathbf{k}\uparrow}\psi_{\mathrm{U},\mathbf{k}\downarrow},\label{H_int_orbitals}
\end{align}
where the label $\Omega$ denotes an orbital configuration which can
be $d\equiv d_{x^{2}-y^{2}}$, $p_{x}$, and $p_{y}$. The on-site
Coulomb interaction strengths for $d_{x^{2}-y^{2}}$ and $p_{x}$
($p_{y}$) orbitals are assumed to be $U_{d}$ and $U_{p}$, respectively,
and the interaction functions are

\begin{align}
\mathcal{V}_{\mathbf{k},\mathbf{k}'}^{(d)}= & v_{11}(\mathbf{k}')v_{11}(-\mathbf{k}')v_{11}^{*}(-\mathbf{k})v_{11}^{*}(\mathbf{k}),\label{Coeff_d}
\end{align}

\begin{align}
\mathcal{V}_{\mathbf{k},\mathbf{k}'}^{(p_{x})}= & v_{21}(\mathbf{k}')v_{21}(-\mathbf{k}')v_{21}^{*}(-\mathbf{k})v_{21}^{*}(\mathbf{k}),\label{Coeff_px}
\end{align}

\begin{align}
\mathcal{V}_{\mathbf{k},\mathbf{k}'}^{(p_{y})}= & v_{31}(\mathbf{k}')v_{31}(-\mathbf{k}')v_{31}^{*}(-\mathbf{k})v_{31}^{*}(\mathbf{k}).\label{Coeff_py}
\end{align}
The functions $v_{ij}$ have been introduced in Eq. (\ref{site_op_interms_band_op}),
and $v_{ij}^{*}$ denotes the complex conjugate of $v_{ij}$. The
interaction Hamiltonian (\ref{H_int}) is obtained as $\hat{H}_{\mathrm{int}}=\hat{H}_{\mathrm{int}}^{(d)}+\hat{H}_{\mathrm{int}}^{(p_{x})}+\hat{H}_{\mathrm{int}}^{(p_{y})}$,
where $\mathcal{V}_{\mathbf{k},\mathbf{k}'}=\mathcal{V}_{\mathbf{k},\mathbf{k}'}^{(d)}+\mathcal{V}_{\mathbf{k},\mathbf{k}'}^{(p_{x})}+\mathcal{V}_{\mathbf{k},\mathbf{k}'}^{(p_{x})}$
and the Fermi sea has been excluded from the first Brillouin zone.

We note that the tight-binding Hamiltonian (\ref{H_tb_1}) in the
basis spanned by the band operators is diagonal. For the submanifold
$\mathcal{S}$ of the upper band, we find the kinetic energy to be

\begin{alignat}{1}
\hat{H}_{\mathrm{kin}}= & \mathop{\sum_{\mathbf{k}\in\,\text{1.BZ}}}_{\sigma\in\{\uparrow,\downarrow\}}E_{\mathbf{k}}^{(\mathrm{U})}\psi_{\mathrm{U},\mathbf{k}\sigma}^{\dagger}\psi_{\mathrm{U},\mathbf{k}\sigma},\label{H_tb in terms of U band}
\end{alignat}
where the Fermi sea will be excluded from the first Brillouin zone
by introducing a chemical potential, $\mu$. Finally, the total Hamiltonian
(\ref{H_tot}) is obtained as $\hat{H}_{\mathrm{tot}}=\hat{H}_{\mathrm{kin}}+\hat{H}_{\mathrm{int}}$.

\setcounter{equation}{0} \renewcommand{\theequation}{D\arabic{equation}}

\section*{appendix d. derivation of the eigenequation (\ref{eigen_Eq_2})}

To derive the pairing equation we calculate the resulting operator
$\hat{H}_{\mathrm{tot}}\left|\Phi\right\rangle $, where $\hat{H}_{\mathrm{tot}}=\hat{H}_{\mathrm{kin}}+\hat{H}_{\mathrm{int}}$
subject to the interacting Fermi sea. For that, first we apply $\hat{H}_{\mathrm{kin}}$
on $\left|\Phi\right\rangle $. The part corresponding to spin-up,
$\hat{H}_{\mathrm{kin}}^{(\uparrow)}$, is obtained to be

\begin{align}
\hat{H}_{\mathrm{kin}}^{(\uparrow)}\left|\Phi\right\rangle = & \sum_{\mathbf{k}\in1.\mathrm{BZ}\setminus\mathrm{FS_{int}}}\xi_{\mathbf{k}}^{(\mathrm{U})}\psi_{\mathrm{U},\mathbf{k}\uparrow}^{\dagger}\psi_{\mathrm{U},\mathbf{k}\uparrow}\sum_{\boldsymbol{\kappa}\in\text{1.BZ}\setminus\mathrm{FS}}\phi(\boldsymbol{\kappa})\nonumber \\
 & \times\psi_{\mathrm{U},\boldsymbol{\kappa}\uparrow}^{\dagger}\psi_{\mathrm{U},-\boldsymbol{\kappa}\downarrow}^{\dagger}\left|\mathrm{FS_{int}}\right\rangle \nonumber \\
= & \sum_{\boldsymbol{\kappa}\in1.\mathrm{BZ}\setminus\mathrm{FS_{int}}}\delta_{\boldsymbol{\kappa}\mathbf{k}}\xi_{\mathbf{k}}^{(\mathrm{U})}\left|\Phi\right\rangle ,\label{H_kin_1_on_Phi}
\end{align}
where $\delta_{\boldsymbol{\kappa}\mathbf{k}}$ denotes the Kronecker
delta. To find the effect of the spin-down part, $\hat{H}_{\mathrm{kin}}^{(\downarrow)}$,
we define $\boldsymbol{\kappa}'\equiv-\boldsymbol{\kappa}$, and rewrite
the singlet-state Cooper pair (\ref{Cooper_pairing_ansatz}) in terms
of $\boldsymbol{\kappa}'$. We obtain that

\begin{equation}
\hat{H}_{\mathrm{kin}}^{(\downarrow)}\left|\Phi\right\rangle =\sum_{\boldsymbol{\kappa}'\in1.\mathrm{BZ}\setminus\mathrm{FS_{int}}}\delta_{\boldsymbol{\kappa}'\mathbf{k}}\xi_{\boldsymbol{\kappa}'}^{(\mathrm{U})}\left|\Phi\right\rangle .\label{H_kin_2_on_Phi}
\end{equation}
Equations (\ref{H_kin_1_on_Phi}) and (\ref{H_kin_2_on_Phi}) result
in:

\begin{align}
\hat{H}_{\mathrm{kin}}\left|\Phi\right\rangle = & \sum_{\mathbf{k}\in\mathrm{1.BZ}\setminus\mathrm{FS_{int}}}\left(\xi_{\mathbf{k}}^{(\mathrm{U})}+\xi_{-\mathbf{k}}^{(\mathrm{U})}\right)\left|\Phi\right\rangle .\label{H_kin_on_Phi}
\end{align}

Next, we apply $\hat{H}_{\mathrm{int}}$ on $\left|\Phi\right\rangle $.
Here we split up the interaction Hamiltonian to the diagonal and off-diagonal
parts. For the diagonal part we obtain:

\begin{widetext}

\begin{align}
\hat{H}_{\mathrm{int}}^{(\mathrm{diag})}\left|\Phi\right\rangle = & \frac{1}{\mathcal{A}}\sum_{\mathbf{k}\in1.\mathrm{BZ}\setminus\mathrm{FS_{int}}}\mathcal{V}_{\mathbf{k},\mathbf{k}}\psi_{\mathrm{U},\mathbf{k}\downarrow}^{\dagger}\psi_{\mathrm{U},-\mathbf{k}\uparrow}^{\dagger}\psi_{\mathrm{U},-\mathbf{k}\uparrow}\psi_{\mathrm{U},\mathbf{k}\downarrow}\sum_{\boldsymbol{\kappa}\in1.\mathrm{BZ}\setminus\mathrm{FS_{int}}}\phi(\boldsymbol{\kappa})\psi_{\mathrm{U},\boldsymbol{\kappa}\uparrow}^{\dagger}\psi_{\mathrm{U},-\boldsymbol{\kappa}\downarrow}^{\dagger}\left|\mathrm{FS_{int}}\right\rangle \nonumber \\
= & \frac{1}{\mathcal{A}}\sum_{\mathbf{k}\in1.\mathrm{BZ}\setminus\mathrm{FS_{int}}}\mathcal{V}_{\mathbf{k},\mathbf{k}}\sum_{\boldsymbol{\kappa}\in1.\mathrm{BZ}\setminus\mathrm{FS_{int}}}\phi(-\mathbf{k})\psi_{\mathrm{U},-\mathbf{k}\uparrow}^{\dagger}\psi_{\mathrm{U},\mathbf{k}\downarrow}^{\dagger}\left|\mathrm{FS_{int}}\right\rangle \nonumber \\
 & +\frac{1}{\mathcal{A}}\sum_{\mathbf{k}\in1.\mathrm{BZ}\setminus\mathrm{FS_{int}}}\mathcal{V}_{\mathbf{k},\mathbf{k}}\sum_{\boldsymbol{\kappa}\in1.\mathrm{BZ}\setminus\mathrm{FS_{int}}}\delta_{\mathbf{k},-\boldsymbol{\kappa}}\phi(\boldsymbol{\kappa})\psi_{\mathrm{U},\mathbf{k}\downarrow}^{\dagger}\psi_{\mathrm{U},-\mathbf{k}\uparrow}\psi_{\mathrm{U},-\mathbf{k}\uparrow}^{\dagger}\psi_{\mathrm{U},\boldsymbol{\kappa}\uparrow}^{\dagger}\left|\mathrm{FS_{int}}\right\rangle \nonumber \\
= & \frac{1}{\mathcal{A}}\sum_{\mathbf{k}\in1.\mathrm{BZ}\setminus\mathrm{FS_{int}}}\mathcal{V}_{\mathbf{k},\mathbf{k}}\left|\Phi\right\rangle ,\label{H_int_diagonal_on_Phi}
\end{align}
\end{widetext}where $\mathcal{A}$ denotes the area of the first
Brillouin zone. For the off-diagonal part we obtain:

\begin{widetext}

\begin{align}
\hat{H}_{\mathrm{int}}^{(\text{off-diag})}\left|\Phi\right\rangle = & \frac{1}{\mathcal{A}}\mathop{\sum_{\mathbf{k},\mathbf{k}'\in1.\mathrm{BZ}\setminus\mathrm{FS_{int}}}}_{\mathbf{k}\neq\mathbf{k}'}\mathcal{V}_{\mathbf{k},\mathbf{k}'}\psi_{\mathrm{U},\mathbf{k}'\downarrow}^{\dagger}\psi_{\mathrm{U},-\mathbf{k}'\uparrow}^{\dagger}\sum_{\boldsymbol{\kappa}\in1.\mathrm{BZ}\setminus\mathrm{FS}}\phi(\boldsymbol{\kappa})\psi_{\mathrm{U},-\mathbf{k}\uparrow}\psi_{\mathrm{U},\mathbf{k}\downarrow}\psi_{\mathrm{U},\boldsymbol{\kappa}\uparrow}^{\dagger}\psi_{\mathrm{U},-\boldsymbol{\kappa}\downarrow}^{\dagger}\left|\mathrm{FS_{int}}\right\rangle \nonumber \\
= & \frac{-1}{\mathcal{A}}\mathop{\sum_{\mathbf{k},\mathbf{k}'\in1.\mathrm{BZ}\setminus\mathrm{FS_{int}}}}_{\mathbf{k}\neq\mathbf{k}'}\mathcal{V}_{\mathbf{k},\mathbf{k}'}\psi_{\mathrm{U},\mathbf{k}'\downarrow}^{\dagger}\psi_{\mathrm{U},-\mathbf{k}'\uparrow}^{\dagger}\sum_{\mathbf{k}'\in1.\mathrm{BZ}\setminus\mathrm{FS_{int}}}\phi(\mathbf{k}')\left(\hat{1}-\psi_{\mathrm{U},\mathbf{k}'\uparrow}^{\dagger}\psi_{\mathrm{U},\mathbf{k}'\uparrow}\right)\left|\mathrm{FS_{int}}\right\rangle \nonumber \\
= & \frac{1}{\mathcal{A}}\mathop{\sum_{\mathbf{k},\mathbf{k}'\in1.\mathrm{BZ}\setminus\mathrm{FS_{int}}}}_{\mathbf{k}\neq\mathbf{k}'}\mathcal{V}_{\mathbf{k},\mathbf{k}'}\left|\Phi\right\rangle .\label{H_int_off_diagonal_on_Phi}
\end{align}
\end{widetext}Equations (\ref{H_int_diagonal_on_Phi}) and (\ref{H_int_off_diagonal_on_Phi})
result in

\begin{align}
\hat{H}_{\mathrm{int}}\left|\Phi\right\rangle = & \frac{1}{\mathcal{A}}\sum_{\mathbf{k},\mathbf{k}'\in1.\mathrm{BZ}\setminus\mathrm{FS_{int}}}\mathcal{V}_{\mathbf{k},\mathbf{k}'}\left|\Phi\right\rangle .\label{H_int_on_Phi}
\end{align}

Finally, we insert Eqs. (\ref{H_kin_on_Phi}) and (\ref{H_int_on_Phi})
into the eigenvalue problem (\ref{eigen_Eq_1}), and arrive at the
pairing equation (\ref{eigen_Eq_2}).

\setcounter{equation}{0} \renewcommand{\theequation}{E\arabic{equation}}

\section*{appendix e. numerical calculation of eq. (\ref{eigen_Eq_2})}

As discussed in the text, to solve Eq. (\ref{eigen_Eq_2}) numerically
we discretize the first Brillouin zone equidistantly following the
relation (\ref{grid_points}). In order to increase the number of
the grid points in each direction and to achieve the numerical stability,
first we calculate the interacting Fermi surface using the relation

\begin{equation}
\mathrm{FS_{\mathrm{int}}=}\Bigl\{\mathbf{k}^{(j)}\in\text{1.BZ}:2E_{\mathbf{k}^{(j)}}^{(\mathrm{U})}+\frac{1}{N^{2}}\mathcal{V}_{\mathbf{k}^{(j)},\mathbf{k}^{(j)}}<2\mu\Bigr\},\label{FS_interacing_discretized}
\end{equation}
and exclude it from the first Brillouin zone. Next, we constitute
the pairing equation (\ref{eigen_Eq_2}) on the reduced momentum space
as

\begin{align}
\left(\xi_{\mathbf{k}_{j}}^{(\mathrm{U})}+\xi_{-\mathbf{k}_{j}}^{(\mathrm{U})}+\frac{1}{N^{2}}\mathcal{V}_{\mathbf{k}_{j},\mathbf{k}_{j}}\right)\phi(\mathbf{k}_{j})\nonumber \\
+\frac{1}{N^{2}}\mathop{\sum_{\mathbf{k}_{j},\mathbf{k}'_{j}\in\mathrm{1.BZ}\setminus\mathrm{FS_{int}}}}_{\mathbf{k}'_{j}\neq\mathbf{k}_{j}}\mathcal{V}_{\mathbf{k}_{j},\mathbf{k}_{j}'}\phi(\mathbf{k}'_{j}) & =\mathcal{E}_{j}\phi(\mathbf{k}_{j}),\label{eigenEq_discretized}
\end{align}
for $j=1,2,\ldots,N$. Finally, we diagonalize Eq. (\ref{eigenEq_discretized}),
and obtain the eigenenergies $\mathcal{E}_{j}$. Among $\mathcal{E}_{j}$,
the desired ground-state energy, $E_{\text{G}},$ is the one which
is negative and has the largest absolute value.

Finally, we notice that the behavior of the desired eigenvalues as
a function of the chemical potential, $\mu$, might display a zigzag
effect due to the finite discretization of the momentum space. To
prevent this behavior, for the noninteracting regime, we calculate
the smallest value of the eigenenergy, $E_{0}$, of Eq. (\ref{eigenEq_discretized})
for the occupied states. Next, for the interacting regime, we add
$E_{0}$ within the first bracket of Eq. (\ref{eigenEq_discretized}),
and calculate the ground-state energy for the unoccupied states.

\setcounter{equation}{0} \renewcommand{\theequation}{F\arabic{equation}}

\section*{appendix f. ground-state solution for the attractive regime}

\begin{figure}[b]
\includegraphics[width=1\columnwidth]{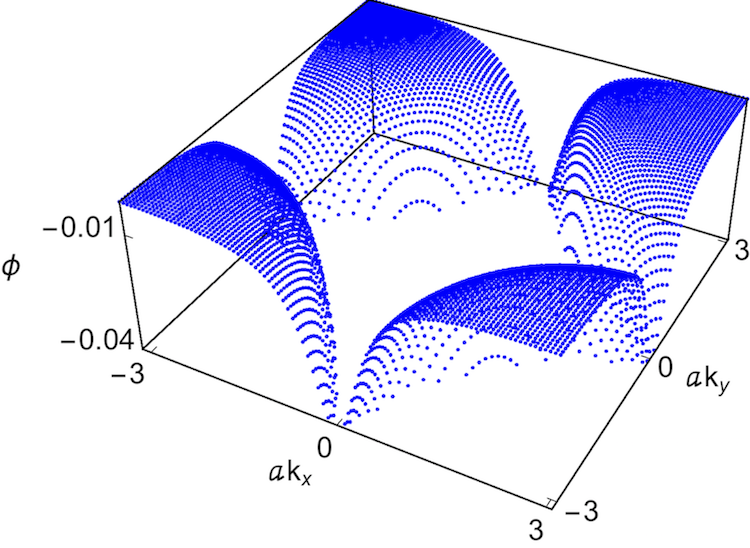}

\caption{Ground-state wave function of the Cooper pair in the attractive regime
of the Fermi-Hubbard model (\ref{H_tot}), where $N=100$, $V_{dp}=3.45\text{ eV}$,
$t_{pd}=1.13\text{ eV}$, $t_{pp}=0.8\text{ eV}$, $\mu\approx-0.679\text{ eV}$,
$U_{d}=-2\text{ eV}$, and $U_{p}=-1\text{ eV}$. The wave function
supports an orbital \emph{s}-wave symmetry.}

\label{Fig7}
\end{figure}

As expected, for the attractive regime, $U_{d},U_{p}<0$, the ground-state
solution supports an orbital \emph{s}-wave symmetry. Figure \ref{Fig7}
shows the wave function for $U_{d}=-2\text{ eV}$ and $U_{p}=-1\text{ eV}.$

\begin{thebibliography}{10}
\bibitem{Lieb_paper}E. H. Lieb, Phys. Rev. Lett. \textbf{62}, 1201
(1989).

\bibitem{Lieb_lattice_Nita}M. Ni{\c t}{\v a}, B. Ostahie, and A.
Aldea, Phys. Rev. B \textbf{87}, 125428 (2013). 

\bibitem{Plakida_HTc_Book}N. Plakida, \emph{High-Temperature Cuprate
Superconductors: Experiment, Theory, and Applications} (Springer,
Berlin, 2010), Chaps. 2 and 3.

\bibitem{Leggett_Cuprates}A. J. Leggett, \emph{Quantum Liquids} (Oxford
University Press, New York, 2006), Chaps. 5 and 7.

\bibitem{BEC_book}C. J. Pethick and H. Smith, \emph{Bose\textendash Einstein
Condensation in Dilute Gases} (Cambridge University Press, New York,
2008), Chap. 5.

\bibitem{trapping_atoms}J. Fort{\'a}gh and C. Zimmermann, Rev. Mod.
Phys. \textbf{79}, 235 (2007).

\bibitem{Lieb_optical_1}R. Shen, L. B. Shao, B. Wang, and D. Y. Xing,
Phys. Rev. B \textbf{81}, 041410(R) (2010). 

\bibitem{Lieb_optical_2}V. Apaja, M. Hyrk{\"a}s, and M. Manninen,
Phys. Rev. A \textbf{82}, 041402(R) (2010).

\bibitem{Lieb_optical_3}S. Taie, H. Ozawa, T. Ichinose, T. Nishio,
S. Nakajima, and Y. Takahashi, Sci. Adv. \textbf{1}, e1500854 (2015).

\bibitem{Lieb_atomic_1}M. R. Slot, T. S. Gardenier, P. H. Jacobse,
G. C. P. van Miert, S. N. Kempkes, S. J. M. Zevenhuizen, C. M. Smith,
D. Vanmaekelbergh, and I. Swart, Nat. Phys. \textbf{13}, 672 (2017).

\bibitem{Lieb_atomic_2}R. Drost, T. Ojanen, A. Harju, and P. Liljeroth,
Nat. Phys. \textbf{13}, 668 (2017).

\bibitem{Lieb_photonic_1}S. Mukherjee, A. Spracklen, D. Choudhury,
N. Goldman, P. {\"O}hberg, E. Andersson, and R. R. Thomson, Phys.
Rev. Lett. \textbf{114}, 245504 (2015).

\bibitem{Lieb_material_1}W. Jiang, H. Huang, and F. Liu, Nat. Comm.
\textbf{10}, 2207 (2019).

\bibitem{Lieb_material_2}B. Cui, X. Zheng, J. Wang, D. Liu, S. Xie,
and B. Huang, Nat. Comm. \textbf{11}, 66 (2020).

\bibitem{Lieb_lattice_Flach}J. D. Bodyfelt, D. Leykam, C. Danieli,
X. Yu, and S. Flach, Phys. Rev. Lett. \textbf{113}, 236403 (2014). 

\bibitem{Flach_flat_band_review}D. Leykam, A. Andreanov, and S. Flach,
Adv. Phys. X \textbf{3}, 1473052 (2018). 

\bibitem{FB_1}N. B. Kopnin, T. T. Heikkil{\"a}, and G. E. Volovik,
Phys. Rev. B \textbf{83}, 220503(R) (2011).

\bibitem{FB_2}V. I. Iglovikov, F. H{\'e}bert, B. Gr{\'e}maud, G.
G. Batrouni, and R. T. Scalettar, Phys. Rev. B \textbf{90}, 094506
(2014). 

\bibitem{FB_3}S. Peotta and P. T{\"o}rm{\"a}, Nat. Comm. \textbf{6},
8944 (2014).

\bibitem{FB_4}A. Julku, S. Peotta, T. I. Vanhala, D. -H. Kim, and
P. T{\"o}rm{\"a}, Phys. Rev. Lett. \textbf{117}, 045303 (2016).

\bibitem{FB_5}K. Kobayashi, M. Okumura, S. Yamada, M. Machida, and
H. Aoki, Phys. Rev. B \textbf{94}, 214501 (2016).

\bibitem{FB_6}M. Tovmasyan, S. Peotta, P. T{\"o}rm{\"a}, and S.
D. Huber, Phys. Rev. B \textbf{94}, 245149 (2016). 

\bibitem{High_Tc_Handbook}J. R. Schrieffer and J. S. Brooks (eds.),
\emph{Handbook of High-Temperature Superconductivity: Theory and Experiment}
(Springer, New York, 2007), Chap. 1.

\bibitem{Lynn_HTc_Book}J. W. Lynn (ed.), \emph{High-Temperature Superconductivity}
(Springer, New York, 1990), Chap. 5.

\bibitem{Anderson_Cuprate_Book}P. W. Anderson, \emph{The Theory of
Superconductivity in the High-}$T_{\mathrm{c}}$ \emph{Cuprates} (Princeton
University Press, Princeton, New Hersey, 1997), Chap. 1.

\bibitem{Uchida_HTc_Book}S. Uchida, \emph{High Temperature Superconductivity:
The Road to Higher Critical Temperature} (Springer, Tokyo, 2015),
Chap. 3.

\bibitem{HTc_Phonon_1}A. A. Abrikosov, Int. J. Mod. Phys. \textbf{13},
3405 (1999).

\bibitem{HTc_Phonon_2}S. R. Park, D. J. Song, C. S. Leem, C. Kim,
C. Kim, B. J. Kim, and H. Eisaki, Phys. Rev. Lett. \textbf{101}, 117006
(2008). 

\bibitem{HTc_Phonon_3}S. Johnston, F. Vernay, B. Moritz, Z. -X. Shen,
N. Nagaosa, J. Zaanen, and T. P. Devereaux, Phys. Rev. B \textbf{82},
064513 (2010).

\bibitem{HTc_Phonon_4}P. J. Carbotte, T. Timusk, and J. Hwang, Rep.
Prog. Phys. \textbf{74}, 066501 (2011). 

\bibitem{HTc_Phonon_5}A. S. Alexandrov, J. H. Samson, and G. Sica,
Europhys. Lett. \textbf{100}, 17011 (2012). 

\bibitem{HTc_Magnon_1}P. Monthoux, A. V. Balatsky, and D. Pines,
Phys. Rev. Lett. \textbf{67}, 3448 (1991).

\bibitem{High_Tc_pairing_mechanism_1}V. M. Krasnov, S. -O. Katterwe,
and A. Rydh, Nat. Comm. \textbf{4}, 2970 (2013). 

\bibitem{HTc_Plasmon_1}H. Rietschel and L. J. Sham, Phys. Rev. B
\textbf{28}, 5100 (1983). 

\bibitem{HTc_Plasmon_2}A. Bill, H. Morawitz, and V. Z. Kresin, Phys.
Rev. B \textbf{68}, 144519 (2003). 

\bibitem{HTc_Plasmon_3}G. S. Atwal and N. W. Ashcroft, Phys. Rev.
B \textbf{70}, 104513 (2004).

\bibitem{HTc_Plasmon_4}E. A. Pashitskii and V. I. Pentegov, Low Temp.
Phys. \textbf{34}, 113 (2008).

\bibitem{Alexandrov_Hubbard_repulsive}A. S. Alexandrov and V. V.
Kabanov, Phys. Rev. Lett. \textbf{106}, 136403 (2011).

\bibitem{Alexandrov_Book}A. S. Alexandrov, \emph{Strong-Coupling
Theory of High-Temperature Superconductivity} (Cambridge University
Press, Cambridge, 2013), Chaps. 6 and 8.

\bibitem{ARPES_1}A. Damascelli, Z. Hussain, and Z. -X. Shen, Rev.
Mod. Phys. \textbf{75}, 473 (2003).

\bibitem{ARPES_2} D. Koralek, J. F. Douglas, N. C. Plumb, Z. Sun,
A. V. Fedorov, M. M. Murnane, H. C. Kapteyn, S. T. Cundiff, Y. Aiura,
K. Oka, H. Eisaki, and D. S. Dessau, Phys. Rev. Lett. \textbf{96},
017005 (2006).

\bibitem{ARPES_3}J. D. Koralek, J. F. Douglas, N. C. Plumb, J. D.
Griffith, S. T. Cundiff, H. C. Kapeteyn, M. M. Murnane, and D. S.
Dessau, Rev. Sci. Instrum. \textbf{78}, 053905 (2007). 

\bibitem{ARPES_4}G. Liu, G. Wang, Y. Zhu, H. Zhang, G. Zhang, X.
Wang, Y. Zhou, W. Zhang, H. Liu, L. Zhao, J. Meng, X. Dong, C. Chen,
Z. Xu, and X. J. Zhou, Rev. Sci. Instrum. \textbf{79}, 023105 (2008). 

\bibitem{ARPES_5}H. Li, X. Zhou, S. Parham, T. J. Reber, H. Berger,
G. B. Arnold, and D. S. Dessau, Nat. Comm. \textbf{9}, 26 (2018).

\bibitem{Starfish}H. Li, X. Zhou, S. Parham, K. N. Gordon, R. D.
Zhong, J. Schneeloch, G. D. Gu, Y. Huang, H. Berger, G. B. Arnold,
and D. S. Dessau, arXiv:1809.02194v2. 

\bibitem{Spalek_HTc_Cuprate}M. Zegrodnik, A. Biborski, M. Fidrysiak,
and J. Spa{\l}ek, Phys. Rev. B \textbf{99}, 104511 (2019).

\bibitem{Spalek_t_J_model} J. Kaczmarczyk, J. B{\"u}nemann, and
J. Spa{\l}ek, New J. Phys. \textbf{16}, 073018 (2014).

\bibitem{Spalek_t_J_U_model} J. Spa{\l}ek, M. Zegrodnik, and J.
Kaczmarczyk, Phys. Rev. B \textbf{95}, 024506 (2017).

\bibitem{H_K_model}P. W. Phillips, L. Yeo, and E. W. Huang, Nat.
Phys. \textbf{16}, 1175 (2020).

\bibitem{DMET_1}G. Knizia and G. K. -L. Chan, Phys. Rev. Lett. \textbf{109},
186404 (2012). 

\bibitem{DMET_2}T. I. Vanhala and P. T{\"o}rm{\"a}, Phys. Rev.
B \textbf{97}, 075112 (2018).

\bibitem{AFQMC}S. Zhang, J. Carlson, and J. E. Gubernatis, Phys.
Rev. Lett. \textbf{74}, 3652 (1995). 

\bibitem{iPEPS}J. Jordan, R. Or{\'u}s, G. Vidal, F. Verstraete,
and J. I. Cirac, Phys. Rev. Lett. \textbf{101}, 250602 (2008). 

\bibitem{DMRG}E. M. Stoudenmire and S. R. White, Annu. Rev. Condens.
Matter Phys. \textbf{3}, 111 (2012).

\bibitem{FLEX_DMFT}M. Kitatani, N. Tsuji, and H. Aoki, Phys. Rev.
B \textbf{95}, 075109 (2017). 

\bibitem{Cooper_paper}L. N. Cooper, Phys. Rev. \textbf{104}, 1189
(1956).

\bibitem{Goodstein_1}D. L. Goodstein, \emph{States of Matter} (Dover,
New York, 1985), Chap. 5.

\bibitem{Abrikosov_book_metals}A. A. Abrikosov, \emph{Fundamentals
of the Theory of Metals} (Dover, New York, 2017), Chap. 16.

\bibitem{Tinkham_1}M. Tinkham, \emph{Introduction to Superconductivity}
(Dover, New York, 2004), Chap. 3.

\bibitem{Aschcroft_Mermin_1}N. W. Ashcroft and N. D. Mermin, \emph{Solid
State Physics} (Brooks/Cole, Belmont, CA, 1976), Chaps. 26 and 34.

\bibitem{Sanayei_1}A. Sanayei, P. Naidon, and L. Mathey Phys. Rev.
Research \textbf{2}, 013341 (2020).

\bibitem{Sanayei_2}A. Sanayei and L. Mathey, arXiv:2007.13511.

\bibitem{hole_doping}To define the hole doping, $\delta_{h}$, here
we follow this convention: At the half-filling the electron density
$n_{e}=1/2$, and the hole doping is vanishing, $\delta_{h}=0$. As
we decrease $n_{e}$, we inject more holes into the desired momentum
space, which increases $\delta_{h}$. Here the hole doping is obtained
as $\delta_{h}=1/2-n_{e}$, for $0\leqslant n_{e}\leqslant1/2$. 

\bibitem{time_of_flight_1}M. Lewenstein, A. Sanpera, and V. Ahufinger,
\emph{Ultracold Atoms in Optical Lattices: Simulating Quantum Many-Body
Systems} (Oxford University Press, 2012), Chap. 14.

\bibitem{time_of_flight_2}M. Inguscio and L. Fallani, \emph{Atomic
Physics: Precise Measurements \& Ultracold Matter} (Oxford University
Press, Oxford, 2013), Chap. 6.

\bibitem{noise_correlation_1}E. Altman, E. Demler, and M. D. Lukin
, Phys. Rev. A \textbf{70}, 013603 (2004).

\bibitem{noise_correlation_2}L. Mathey, A. Vishwanath, and E. Altman,
Phys. Rev. A \textbf{79}, 013609 (2009). 

\bibitem{noise_correlation_3}V. Pal Singh, W. Weimer, K. Morgener,
J. Siegl, K. Hueck, N. Luick, H. Moritz, and L. Mathey Phys. Rev.
A \textbf{93}, 023634 (2016).

\bibitem{cubic_eq}I. J. Zucker, Math. Gazet. \textbf{92}, 264 (2008).

\bibitem{momentum_transfer}By \textquotedblleft momentum transfer\textquotedblright{}
we mean the difference of the in-state and out-state momenta of a
particle; see, e.g., J. R. Taylor, \emph{Scattering Theory: The Quantum
Theory of Nonrelativistic Collisions} (Dover, New York, 2006).
\end{thebibliography}
\end{document}